\documentclass[12pt]{article}
\usepackage{graphicx}
\usepackage{authblk}

\def\be{\begin{equation}}
\def\ee{\end{equation}}
\def\bea{\begin{eqnarray}}
\def\eea{\end{eqnarray}}

\title{Effects of plasma on physical properties of water: nanocrystalline-to-amorphous phase transition and improving produce washing}

\author{Jinjie He$^1$, Alexander Rabinovich$^1$, Dmitri Vainchtein$^1$, Alexander Fridman}
\affil{Nyheim Plasma Institute, Drexel University, Camden, NJ, US}

\author{Christopher Sales}
\affil{Civil, Architectural and Environmental Engineering, Drexel University, Philadelphia, PA, USA}

\author{Mikhail N. Shneider}
\affil{Department of Mechanical and Aerospace Engineering, Princeton University, Princeton, NJ, USA}

\begin{document}

\maketitle

\begin{abstract}

Recently is was discovered in various applications that many physical and chemical properties of water change their temperature dependence between about $35$ and $60$ degrees Celsius. In particular, heat conductance, light absorption, and surface tension all change their temperature dependence. These drastic changes were associated with water gradually changing its mesoscopic structure: while at the higher temperatures water is a uniform media (amorphous state), at the temperatures below transition it consists of many nano-to-micro-scale clusters (crystalline state). This transition is similar to the second order phase transition. In the present paper we show that treating water with non-thermal plasma (adding plasma-created active compounds) can lower the temperature of the transition and thus cause a significant change in such physical quantities as surface tension, viscosity, freezing rate, and wettability and washability. We present analytical estimates of the transition temperature shift based on the Debye–Hückel theory. We discuss the produce-washing experiments that illustrate the predicted effects. 

\end{abstract}

\section{Introduction}

With water being of the most abundant elements on Earth, its physical properties play key roles in many very diverse properties. Recently it was observed that otherwise smooth temperature dependence of many of these properties undergo a localized, change between about $35$ and $60$ degrees Celsius. In particular, electrical conductance, absorption coefficient, refractive index, thermal conductivity, and surface tension all undergo a jump in their temperature derivative (see \cite{Vargaftik:1983,Rusuniak:2004,Shneider:2019} and review \cite{Maestro:2016}). Each physical property has its own transition temperature. It was suggested in several publications that this transition is related to the change of the internal structure of water (see, e.g., \cite{Yakhno:2019,Pettersson:2016,Brites:2020} and references therein). In particular, the structural changes were shown experimentally using the X-ray absorption spectroscopy (XAS) \cite{Wernet:2004}, X-ray emission spectroscopy (XES) \cite{Tokushima:2008}, vibrational spectroscopy \cite{Nilsson:2015} (see also a review \cite{Pettersson:2016}). Changes in the derivatives in such quantities as thermal conductivity is a telltale sign of second order phase transition (similar to the Ferromagnet-Paramagnet second order phase transition). Above the transition range water is essentially isotropic media of separate molecules (also called amorphous state). Below the transition range, most of the molecules are in the form of clusters, with molecules being hold in place by the Hydrogen bonds. These clusters vary in sizes with the distribution strongly depending on the temperature. Typical sizes of the clusters were estimated to be on the scale of tens of nanometers. 

In the present paper we show that adding plasma-activated water changes (decreases) the temperature of the transition. At the room temperature this effects manifests itself in that the physical properties of water start more resembling the properties of the amorphous state (for example, lower surface tension and viscosity). The structure of this paper is as follows. In Section 2 we recall equilibrium between the two states using the notion of free energy. In Section 3 we follow the Debye–Hückel theory to introduce the additional term due to plasma. In Section 4 we discuss the changes of physical properties of plasma-activated water. In Section 5 we present the experiments that illustrate and confirm changes in wettability and washability of plasma-activated water. Section 6 contains discussion and conclusions. 

\section{Free energy and the phase transition}

\begin{figure}[t]
    \centering
    \includegraphics[width=0.9\textwidth]{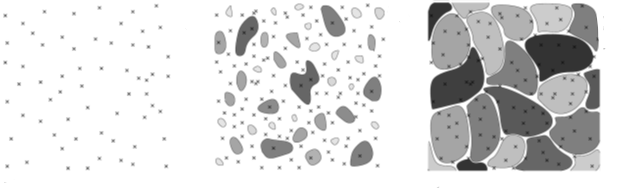}
    \caption{Schematic view of the phase transition from the amorphous (left) to fully-crystalline (right) state (from \cite{Tanaka:1999a}).}
    \label{fig:11}
\end{figure}

It was suggested in \cite{Tanaka:1999a,Tanaka:1999b,Tanaka:2000} that the phase transition between the crystalline (also called structured) and the amorphous (also called random) states of water (see Fig.\ref{fig:11}) can be described in terms of the free energy, which can be written as 
\bea
F &=& U - T \sigma + q \left( q V_S + \left( 1 -q \right) V_r \right) \nonumber \\
&=& q E_s + \left( 1 -q \right) E_r + \left( q V_S + \left( 1 -q \right) V_r \right) P 
\label{eq1} \\
&+& k_B T \left( q \ln\frac{q}{g_s} + \left( 1 -q \right) \ln\frac{1-q}{g_r} \right)
\nonumber 
\eea
In (\ref{eq1}), $F$ is a Free energy, $\sigma$ is the entropy, $q$ is a percentage of structured state, $E_{s,r}$ are the specific energies of the two states $(E_s < E_r)$, $E_{s,r}$ are the specific volumes, $g_{s,r}$ are the statistical degeneracy, $(g_s \ll g_r)$, $T$ is temperature, and $k_B$ is the Boltzmann's constant (see \cite{Tanaka:2000}). The value of $q$ play the role of the order parameter: the equilibrium value of $q$, defined by the minimum of Free energy at a given temperature, defines the relative abundance of the two states. 

Based on (\ref{eq1}), the equilibrium value of $q$ is 

\be
\frac{q_{eq}}{1-q_{eq}} = \frac{g_s}{g_r} \exp\left( -\frac{E_s - E_r}{k_B T} - \left(V_S - V_r \right) P \right)
\label{eq2}  
\ee
In (\ref{eq2}), the difference in the specific energy is $\Delta E = E_s - E_r \approx 1800^\circ$K, \cite{Tanaka:2000}. 

Besides just the ratio of the abundances of the two states, the value of $q$ defines the distribution of sizes of the nano-crystals, constituting the structured state of water. 

\section{Changes of the Free energy in the presence of plasma. The Debye–Huckel term. }

When ions are added to a gas or fluid, the Free Energy changes. Free energy in the presence of ions contains an additional, Debye–Huckel, term:
\be
u_{DH} = - \sum_{i=1}^M  \frac{N_i z_i^2}2 \frac{e^2 \kappa}{4 \pi \epsilon_r \epsilon_0} \frac{1}{1+ \kappa a_i}, \quad \kappa^2 = \frac{2 e^2}{\epsilon_r \epsilon_0 k_B T} \sum_{i=1}^M z_i^2 n_i
\label{eq3}  
\ee
where $i$ denotes the type of ion specie, $N_i$ is the ion's concentration, $a_i$ is the ion's radius,  $z_i$ is the charge of an ionic specie, $M$ is the total number of the ionic species, $\kappa$ is the inverse of the Debye screening length. The ions are unlikely to penetrate into the clusters, thus in the first approximation their impact is proportional to the percentage of water in the amorphous phase. Thus we get  
\be
F_{DH} = F + (1-q)u_{DH} 
\label{eq4}  
\ee
While the Debye–Huckel theory is only approximate, it can provide an qualitative description of in what direction the addition of plasma changes the transition temperature. Form the very fact that the Debye–Huckel term 'favors' the uniform (amorphous) state, it is clear that the transition temperature decreases.

\section{Changes of physical properties of water in the presence of plasma}

There are several major differences between the properties of the Plasma-activated water (PAW) and the pure (or tab) water. However, recently the studies of PAW concentrate almost entirely on the changes of chemical properties, which, in particular, are related to the presence of reactive Oxygen and Nitrogen reactive species (ROS and RNS) (see \cite{Fridman:2008, Graves:2012} and references therein). Meanwhile, changes of physical properties were almost not discussed (see \cite{Fridman_Kennedy:2008}). The estimates based on the Debye–Huckel theory indicate that addition of plasma should change the very structure of water, the balance between structured and random states. Adding plasma to water not only changes (reduces) the phase transition temperature between the random and structured states. At the room temperature the ``pure'' water is solidly in the structured state, but adding plasma reduces the effective value of the order parameter $q$. Consequently, both the number of the clusters and the typical sizes of the clusters are reduced, (see \cite{Brites:2020}). 

One of the consequences of changing the order parameter is that it changes the surface tension. Experiments reported in \cite{Vargaftik:1983,Rusuniak:2004} (see also review \cite{Maestro:2016} and references therein) showed that reducing the amount and size of clusters reduces the surface tension of water. In the everyday life, there are two standard approaches to reduce surface tension: increasing the temperature and/or adding a surfactant (such as soap). Lowering surface tension improves the washing effectiveness of water: the smaller is the surface tension, the easier it is for water to pick a generalised ``dirt,'' or anything else, from a surface of a washed object and to remove it. 

\section{Improved cleaning of the produce by PAW}

We performed two sets of experiments to verify that treating water by plasma at room temperature changes physical properties of water. Specifically, we were interested to see how the plasma treatment changes the wash-out ability of water: the efficiency of the bacterial removal.

Currently there are two approaches (technologies) to surface cleaning and/or sterilization: (1) Disinfection –- when water inactivates microorganisms, possibly leaving them in place, and (2) Wash-out –- when water detaches microorganisms from the surface and subsequently disposes them. Until recently, all studies about PAW focused on (chemical) inactivation. The main difficulties of inactivation on leafy produce are due to two factors. Complicated surfaces with folds make it hard for water to reach the microorganisms that hide in pores. Additionally, PAW reacts with organic load in water. As it was mentioned above, plasma stimulated water transition from nano crystal to amorphous phase results in decrease of water viscosity, surface tension and increase of surfactancy, which finally leads to significant increase of washability. This effect of plasma stimulated washability enhancement can be demonstrated by experiments with washing of fresh produce in plasma activated water PAW. While it is well known that PAW effectively suppress pathogens on the fresh produce surface, we demonstrate that PAW also stimulate removing (washing out) of these microorganisms from the produce surface with further killing of them already in liquid, \cite{Fridman_Friedman:2013, Kim:2013}.

\subsection{Plasma production}
\begin{figure}[t]
    \centering
    \includegraphics[width=0.5\textwidth]{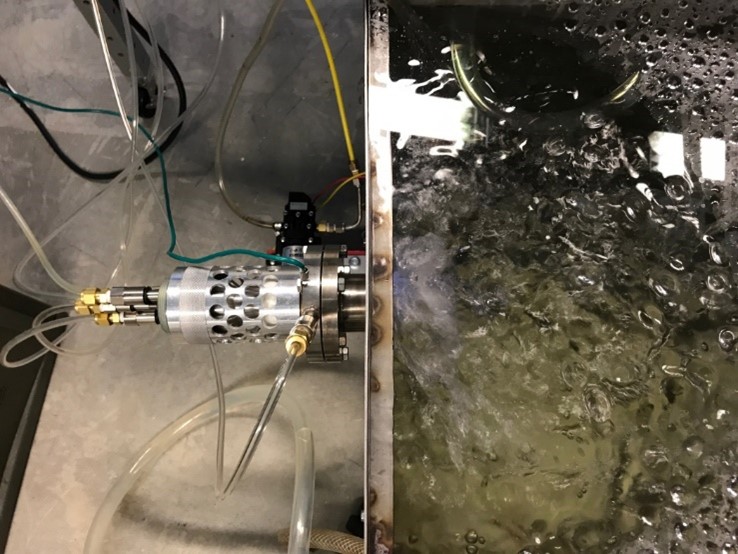}
    \caption{Operation of gliding arc plasmatron in a water tank during PAW production.}
    \label{fig:1}
\end{figure}

In our experiments we used gliding arc plasma. This transient type of discharge is non-equilibrium with a relatively high microarc temperature (about $1600-1800^\circ$K). This setup is optimal for plasma-activated water production. The system operation is illustrated in Fig.~\ref{fig:1}. Gliding arc plasmatron is connected to a tank filled with tap water. Air is injected tangentially into the gap between two cylindrical electrodes and creates a vortex. Power supply applies a voltage between the high voltage electrode and the ground electrode. Plasma discharge starts between two electrodes. The air vortex stretches and rotates the gliding arc and produces the plasma zone inside the plasmatron. Water is injected by a water pump into the plasmatron, passes through the plasma zone, and collected at the exit of the plasma system. After being processed in the gliding arc plasmatron, water gains sterilization properties due to production in plasma different kinds of active species such as OH radicals, hydrogen peroxide, NOx, etc. During plasmatron operation the air that coming out of plasma zone reacts with the tap water producing plasma activated water PAW. Usually pH of PAW is $3-3.5$ compared to $6-6.5$ of tap water.

System parameters are as follows. The water flow rate was $60$mL/min, the plasma air flow rate was $50$ SLPM, the wall protection air was $24$ SLPM, and the water atomization air was $51$ SLPM. The experiment was performed using $1100$W plasma power. After PAW was generated, a chiller was used to cool the water down to $3-5^\circ$C. Temperature, conductivity, and pH were measured by a pH meter after cooling. The abundance of NO3-, NO2-, and H2O2 was measured by test strips. 

\subsection{Bacteria preparation}
In the current experimental setup we used Rifampicin resistant {\it E. coli O157: H7} (ATCC 700728). Fresh cultured {\it E.coli} was incubated in Tryptone Soy Broth (Becton Dickinson) containing $20$mg/L Rifampicin (Sigma Aldrich) at $37^\circ$C overnight. The concentration of {\it E.coli} was approximately $8-9$ log CFU/mL. To prepare the inoculum, $10$ mL of fresh cultured {\it E.coli} was centrifuged at $3000$ rcf for $15$ min. Pellets were re-suspended in 10 mL of sterile phosphate-buffered saline (PBS). Then bacterial suspension was centrifuged and re-suspended again to remove extra organic in culture broth. The bacterial suspension was then diluted 1:10 to obtain a concentration of $7-8$ log CFU/mL.

\subsection{Produce preparation}
For the experiments with fresh produce washing we used Romain Lettuce, obtained from by Dole produce company. Each leaf was washed twice by sterile phosphate-buffered saline (PBS) to remove extra organic in culture broth. After washing, the lettuce was cut into pieces, each weighting approximately $1$ gram. After that, $100 \mu$L of {\it E.coli} was spot-inoculated on each piece. The amount of {\it e.coli} on romaine was measured to be $104-105$ CFU/g. Inoculated romaine was stored in unsealed plastic bags and kept at $4^\circ$ C for $20$ hours to allow {\it e.coli} attach to the surface of leaves. E.coli was quantified by plate count.

\subsection{Washing Procedure}
Plasma-activated water (PAW) was collected and cooled in a chiller to $4^\circ$ C before washing. $10$ grams of lettuce that was inoculated with {\it E.coli} was split into two teabags and then dipped up and down in $200$mL of PAW for $1$min. After wash, lettuce was de-watered in a salad spinner and transferred into a stomacher bag with $10$mL of sterile PBS and homogenized with a Seward 80 Circulator stomacher (Seward Laboratory Systems Inc.) at a normal speed for $120$ seconds. The obtained produce slurry was serially diluted with PBS and cultured at Rifampicin-contained TSA medium at $37^\circ$C overnight to quantify viable {\it E. coli} remained on the lettuce after wash. The decontamination efficiency was computed as 
\be
\mbox{Decontamination efficiency} = \log_{10} \left(A_0/A \right)
\label{eq5}  
\ee
Where $A_0$ is the concentration of {\it E. coli} on inoculated but untreated group and $A$ is the concentration of {\it E. coli} that remained on produce after washing.

\subsection{Quantification of remained DNA}
The DNA of {\it E. coli} that remained on lettuce after wash was extracted from the E.coli inoculated pieces of lettuce via QIAamp Fast DNA Stool Mini Kit (Qiagen). The qPCR primer sets targeting {\it E. coli O157:H7} (F 5’   TAAATGGCACCTGCAACGGA - 3’; R 5’ - GTCATCTTACGGCTGCGGAT- 3’) was ordered from IDT DNA and used for qPCR analysis. A fast SYBR Green qPCR assay was applied to obtain the concentration of {\it E. coli}.

The QuantStudio 3 Real-Time PCR System and Applied Biosystems Fast SYBR Green Master Mix were used to conduct all qPCR assays. Total qPCR reaction volume was $20 \mu$L. Each reaction mixture contained $6 \mu$M of the forward and reverse primers, $2 \mu$L of template DNA and $10 \mu$L of fast SYBR Green Master Mix. The program employed: pre-incubation for $20$s at $95^\circ$C; $40$ amplification cycles of $1$s of denaturing at $95^\circ$C and $20$s of annealing at $60^\circ$C; and, finally, $1$s of $95^\circ$C, $20$s of $60^\circ$C and $1$s of $95^\circ$°C for melt curve. All assays were conducted in triplicates of each sample with negative controls and positive controls.

\subsection{Results}
The decontamination efficiency was determined by comparing the plate count of washed and unwashed samples. Therefore, the decontamination of lettuce washed in water includes inactivation of {\it E. coli} on the surface of lettuce and wash-out to the water.  The washed lettuce showed 0.7-log reduction in PAW compared to unwashed lettuce, while the decontamination efficiency was 0.5-log reduction when washed in tap water (Table 1). So PAW has a stronger ability to decontaminate {\it E. coli} on lettuce.

The total amount of DNA of {\it E. coli} that remained on the lettuce after washing in PAW or tap water was measured by qPCR. This result included both alive and dead {\it E. coli} on the surface of lettuce. More total DNA was found after washing in tap water (105490) than after washing in PAW (69757), which means more {\it E. coli} (either alive or dead) was washed out from the surface of lettuce. 

Therefore it was demonstrated that PAW not only kills pathogens both on the produce surface and washing water, but also enhance washing ability of the water, as a result of water transition from nanocrystalline to amorphous state. The total remaining amount of {\it E. coli} (dead or alive) is significantly lower for the PAW compared with tap water, which cannot be explained by the mere presence of ROS or RNS. 

\begin{table}[]
    \centering
    \begin{tabular}{|c|c|c|}
         &  Wash out efficiency [log] & {\it E. coli} remained [\%] \\
        PAW & 0.67 $\pm$ 0.21	& 24\% $\pm$ 11 \% \\
PAW w/organic load &	0.70 $\pm$ 0.33 & 26\% $\pm$ 19\% \\
Tap & 0.55 $\pm$ 0.32 &	36\% $\pm$ 25 \%
    \end{tabular}
    \caption{The effectiveness of PAW in bacterial removal. }
    \label{t1}
\end{table}

\section{Discussion and conclusions}

In the present paper we showed that plasma-activated water had a lower temperature of the amorphous-crystalline transition: rather than being in the range of $40-60^\circ$C. it moves closer to the room temperature. As a result, at a given (room) temperature this shift makes water to behave more resembling the amorphous state, which is characterised by, in particular, lower surface temperature and lower viscosity. Both these effects have significant impact on the properties and effectiveness of washing using the PAW. Reduced surface tension ease picking of dirt and bacteria from the surfaces, as to be taken from the surface into the balk of water requires overcoming of the potential barrier created by the surface tension. Meanwhile, reducing viscosity makes it easier for water to enter narrow channels and cracks on the surface of the produce, that are known locations of the bacteria. 

The surface tension and viscosity are not the only quantities that feels the effect of adding plasma. Preliminary experiments indicate that a lot of kinetic quantities also change. Among them are light absorption, and thermal conductivity. 

\section*{Acknowledgements}
This material is based upon work partially supported by the Princeton Collaborative
Research Facility (PCRF) and supported by the U.S. Department of Energy (DOE) under Contract No. DE-SC0021378; and the Center for Produce Safety under Award 2020CPS05. 

\section*{Declaration of originality} 
The article has not been published elsewhere and has not been simultaneously submitted for publication elsewhere. The original electronic files of drawings, photos and the article will be retained by the authors until the conclusion of the publication process. We confirm that all tables and figures are your original work and no permissions are required.



\end{document}